\documentclass[preprint,12pt]{elsarticle}

\usepackage{amssymb}
\usepackage{multirow}
\usepackage{booktabs}
\usepackage{amssymb}
\usepackage{mathrsfs}
\usepackage{longtable}
\usepackage{lscape}
\usepackage{tabularx}
\usepackage{epsfig}
\usepackage{colortbl}
\usepackage{subfigure}
\usepackage{graphicx}
\journal{Physica A}

\begin{document}

\begin{frontmatter}



\title{Tsallis entropy of complex networks}


\author[swu]{Qi Zhang}
\author[swu]{Meizhu Li}
\author[swu,NWPU,vu]{Yong Deng\corref{cor}}
\ead{ydeng@swu.edu.cn, prof.deng@hotmail.com}
\author[vu]{Sankaran Mahadevan}

\cortext[cor]{Corresponding author: Yong Deng, School of Computer and Information Science, Southwest University, Chongqing, 400715, China.}

\address[swu]{School of Computer and Information Science, Southwest University, Chongqing, 400715, China}
\address[NWPU]{School of Automation, Northwestern Polytechnical University, Xian, Shaanxi 710072, China}
\address[vu]{School of Engineering, Vanderbilt University, Nashville, TN, 37235, USA}

\begin{abstract}
    How $complex$ of the complex networks has attracted many researchers to explore it. The entropy is an useful method to describe the degree of the $complex$ of the complex networks. In this paper, a new method which is based on the Tsallis entropy is proposed to describe the $complex$ of the complex networks. The results in this paper show that the $complex$ of the complex networks not only decided by the structure property of the complex networks, but also influenced by the relationship between each nodes. In other word, which kinds of nodes are chosen as the main part of the complex networks will influence the value of the entropy of the complex networks. The value of $q$ in the Tsallis entropy of the complex networks is used to decided which kinds of nodes will be chosen as the main part in the complex networks. The proposed Tsallis entropy of the complex networks is a generalised method to describe the property of the complex networks.
\end{abstract}
\end{frontmatter}

\section{Introduction}
\label{Introduction}
The complex networks is a system which composed of many interacting parts. There are many real systems can be modeled as the complex ntworks, such as the biological, social and technological systems \cite{albert2000error,newman2003structure}. Many property of the complex networks have illuminated by these researchers in this filed, such as the network topology and dynamics \cite{watts1998collective,newman2006structure}, the property of the network structure \cite{newman2003structure,barthelemy2004betweenness}, the self-similarity and fractal property of the complex networks\cite{song2005self,wei2014informationdimension}, the controllability and the synchronization of the complex networks \cite{liu2011controllability,arenas2008synchronization} and so on \cite{barrat2004architecture,barabasi2009scale,barabasi1999emergence,barabasi2009scale,song2005self,teixeira2010complex}.  Among these researches, one of the most important topic is how to describe the $complex$ of the complex networks. Depends on the information theory, the entropy of the complex networks is proposed to measure the $complex$ of the complex networks by some researchers \cite{bianconi2008entropy,anand2009entropy,xiao2008symmetry}. Most of the existing methods in this filed are based on the shannon entropy. In this paper, a new entropy of the complex networks is proposed based on the Tsallis entropy.

In the existing method, the $complex$ of the complex networks is decided by the structure property, such as the connection between each node, the betweenness centrality and the clustering coefficient. However, in the Tsallis entropy the $complex$ of the complex networks also decided by the relationship between each nodes. If these nodes with small value of degree are chosen as the main construction of the complex networks, then the network become more $complex$, else when these nodes with big value of degree are chosen as the main construction of the complex networks, then the networks become more orderly. The value of $q$ in the Tsallis entropy of the complex networks decides which kinds of nodes are chosen as the main construction of the complex networks. When the value of $q$ is smaller than 1, the nodes with small value of degree are chosen as the main construction of the complex networks and the network become more $complex$. The value of entropy is increase. When the value of $q$ is bigger than 1, the nodes with big value of degree are chosen as the main construction of the complex networks and the network become more orderly. The value of entropy is decrease. When the value of $q$ is equal to 1, the Tsallis entropy of the complex networks is degenerated to the degree structure entropy of the complex networks. The $complex$ of the complex networks is decided by the structure property. When the value of $q$ is equal to 0, each node has the same influence on the whole network. The value of the entropy of the complex networks reach to the maximum and the network become the most $complex$.

It clear that the Tsallis entropy of the complex networks is a generalised method to describe the property of the complex networks.

The rest of this paper is organised as follows. Section \ref{Rreparatorywork} introduces some preliminaries of this work. In section \ref{Tsallis_E_C_N}, a new entropy of the complex networks based on the Tsallis entropy is proposed. The application of the proposed method is illustrated in section \ref{application}. Conclusion is given in Section \ref{conclusion}.
\section{Preliminaries}
\label{Rreparatorywork}
In this section, the definitions of the degree centrality and the Tsallis entropy are shown as follows.
\subsection{Degree centrality}
\label{Degree}
The degree of one node in a network is the number of the edges connected to the node. Most of the properties of the complex network are based on the degree distribution, such as the clustering coefficient, the community structure and so on. In the network, ${Degree(i)}$ represents the degree of the $i$th node. To calculate the structure entropy of the complex networks, the degree of the complex networks is defined as follows \cite{xiao2008symmetry,xu2013degree}:

\begin{equation}\label{KI}
{d_i} = \frac{{{Degree(i)}}}{{\sum\limits_{i = 1}^n {{Degree(i)}} }}
\end{equation}

Where the ${d_i}$ represents the new degree of the $i$th node which will be used in the calculation of entropy \cite{newman2003structure}, $n$ represents the total numbers of the nodes in the whole networks.
\subsection{Tsallis entropy}
\label{Tsallis entropy}
The entropy is defined by Clausius for thermodynamics \cite{clausius1867mechanical}. For a finite discrete set of probabilities, the definition of the Boltzmann-Gibbs entropy \cite{gibbs2010elementary} is given as follows:

\begin{equation}\label{S_BG}
{S_{BG}} =  -k \sum\limits_{i = 1}^N {{p_i}} \ln {p_i}
\end{equation}

The conventional constant $k$ is the Boltzmann universal constant for thermodynamic systems. The value of $k$ will be taken to be unity in information theory \cite{tsallis2010nonadditive,shannon2001mathematical}.

In 1988, a more general form for entropy have been proposed by Tsallis  \cite{tsallis1988possible}. It is shown as follows:

\begin{equation}\label{S_q}
{S_q} =  - k\sum\limits_{i = 1}^N {{p_i}} {\ln _q}\frac{1}{{{p_i}}}
\end{equation}

The $q-logarithmic$ function in the Eq. (\ref{S_q}) is presented as follows \cite{tsallis2010nonadditive}:
\begin{equation}\label{ln_q}
{\ln _q}{p_i} = \frac{{{p_i}^{1 - q} - 1}}{{1 - q}}({p_i} > 0;q \in \Re ;l{n_1}{p_i} = ln{p_i})
\end{equation}

Based on the Eq. (\ref{ln_q}), the Eq. (\ref{S_q}) can be rewritten as follows:

\begin{equation}\label{S_q1}
{S_q} =  - k\sum\limits_{i = 1}^N {{p_i}} \frac{{{p_i}^{q - 1} - 1}}{{1 - q}}
\end{equation}

\begin{equation}\label{S_q2}
{S_q} =  - k\sum\limits_{i = 1}^N {\frac{{{p_i}^q - {p_i}}}{{1 - q}}}
\end{equation}

\begin{equation}\label{S_q1}
{S_q} = k\frac{{1 - \sum\limits_{i = 1}^N {{p_i}^q} }}{{q - 1}}
\end{equation}

Where ${N}$ is the number of the subsystems.

Based on the Tsallis entropy, the nonextensive theory is established by Tsallis et.al \cite{tsallis2010nonadditive}.
\section{Tsallis entropy of the complex networks}
\label{Tsallis_E_C_N}
The Tsallis entropy is the foundation of the nonextensive statistical mechanic. In the complex networks, the basic construction in the structure property is the node's degree. Therefore, in this paper, each node's degree is used as the basic factor of the entropy. In order to use the degree in the calculation of the entropy, the form of each node's degree is transformed in the probability. The details are shown in the Eq.\ref{KI}. Depends on the definition of the Tsallis entropy, the Tsallis entropy of the complex networks is defined as follows.

\begin{equation}\label{S_di}
{S_q}' = k\frac{{1 - \sum\limits_{i = 1}^n {{d_i}^q} }}{{q - 1}}
\end{equation}

Where the ${S_q}'$ represents the Tsallis entropy of the complex networks, the ${d_i}$ represents the new degree of the $i$th node, the $n$ is the total number of the nodes in the whole network. The $q$ is the entropic index in the Tsallis entropy which is used to describe the nonextensive additivity in the complex networks.

\begin{figure}[htbp]
  \centering
  \includegraphics[scale=0.5]{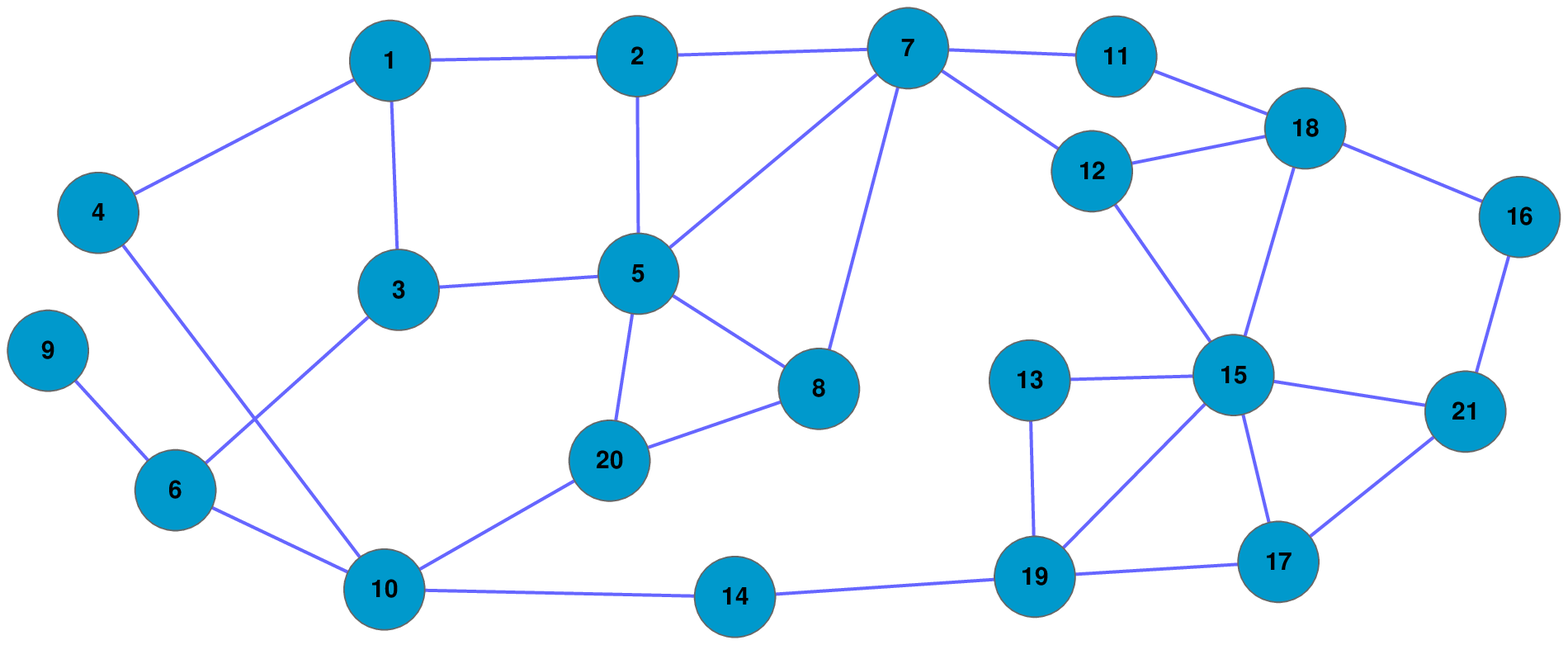}\\
  \caption{The example network.}\label{E_N}
\end{figure}

In order to illuminate the principle of the Tsallis entropy of the complex networks, a example network is shown in the Fig.(\ref{E_N}).

The details of the degree distribution of the example network is shown in the Table \ref{tab:eN_d}.
\begin{table}[htbp]
\tiny
\addtolength{\tabcolsep}{-5pt}
  \centering
  \caption{The degree distribution of the example network}
    \begin{tabular}{cccccccccccccccccccccc}
    \toprule
    node number & 1     & 2     & 3     & 4     & 5     & 6     & 7     & 8     & 9     & 10    & 11    & 12    & 13    & 14    & 15    & 16    & 17    & 18    & 19    & 20    & 21 \\
    \midrule
    $Degree$ & 3     & 3     & 3     & 2     & 5     & 3     & 5     & 3     & 1     & 4     & 2     & 3     & 2     & 2     & 6     & 2     & 3     & 4     & 4     & 3     & 3 \\
    $d_i$   & 0.045 & 0.045 & 0.045 & 0.03 & 0.075 & 0.045 & 0.075 & 0.045 & 0.015 & 0.06& 0.03 & 0.045 & 0.03 & 0.03 & 0.09 & 0.03 & 0.045 & 0.06 & 0.06& 0.045& 0.045 \\
    \bottomrule
    \end{tabular}%
  \label{tab:eN_d}%
\end{table}%

 The proposed method is used to calculate the entropy of the example network. The details are shown in the Table \ref{tab:EN_E}.

\begin{table}[htbp]
\tiny
\addtolength{\tabcolsep}{-3pt}
  \centering
  \caption{The Tsallis entropy of the example network}
    \begin{tabular}{ccccccccccc}
    \toprule
      The value of q   & q=0.5 & \textbf{q=1 }  & q=1.5 & q=2.0 & q=2.5 & q=3.0 & q=3.5 & q=4.0 & q=4.5 & q=5.0 \\
    \midrule
   ${S_q}'$   & 7.005657 & \textbf{2.976405} & 1.541335 & 0.945822 & 0.657896 & 0.498362 & 0.399666 & 0.333261 & 0.285698 & 0.249996 \\
    \bottomrule
    \end{tabular}%
  \label{tab:EN_E}%
\end{table}%

When the value of $q$ is equal to 1, the Tsallis entropy is degenerated to the degree entropy \cite{xu2013degree}. It is clear that follow the increase of the value of $q$, the value of the Tsallis entropy of the complex networks is decrease. Based on the definition of the entropy, the bigger the value of the entropy, the more $complex$ of the network.
From the definition of the Tsallis entropy, the value of the entropic index $q$ can be used to change the construction of the entropy. In other word, the value of $q$ represents the relationship among those nodes. Combine with the complex networks, the influence of each node's degree on the entropy is changed by the value of $q$. The relationship between the value of $q$ and the entropy of the complex networks show as follows:

\begin{description}
  \item[Case 1] \textbf{When} ${q=0}$, each node has the same influence on the whole network. The value of the Tsallis entropy of the complex networks equal to the maximume. The networks become the most $complex$.
  \item[Case 2] \textbf{When} ${q \to 1}$, these nodes with small value of degree paly important role in the construction of the entropy, they are chosen as the main construction of the complex networks.
  \item[Case 3] \textbf{When} ${q=1}$, the influence of each node on the network is based on the value of degree for each node. The Tsallis entropy of the complex networks degenerates to the degree structure entropy. The structure property of the complex networks decide the degree of the $complex$ of the complex networks.
  \item[Case 4] \textbf{When} ${q \to \infty}$, there nodes with big value of degree play important role in the construction of the entropy, they are chosen as the main construction of the complex networks. The value of the Tsallis entropy is tended to 0. The complex networks is tended to orderly.
\end{description}

According to the definition of the Tsallis entropy of the complex networks, the value of the entropic index $q$ is used to describe the different relationship among these nodes. When the value of $q$ is smaller than 1, the nodes with small value of degree are important than the nodes with big value of degree. The edges among those nodes with small value of degree become the main part of the complex network. Because these nodes with small value of degree is the majority in the complex networks so the whole network become more complex. When the value of $q$ is equal to 0, the nodes in the networks is equal to each other, the value of the entropy reach to the maximume. When the value of $q$ is equal to 1, the Tsallis entropy is degenerated to the degree structure entropy, the degree of the $complex$ for the complex networks is decided by the structure property. In other word, the complex of the complex networks is decided by the degree distribution. When the value of $q$ is trended to the $\infty$ the construction of the complex networks is decided by the node which has a biggest value of degree, the value of the entropy of the complex networks is equal to 0, and the complex networks is orderly.

The result shows that, the $complex$ of the complex networks is not only decided by the structure of the complex network, but also influence by the kind of the relationship between each node. The Tsallis entropy of the complex networks is more generalised method to describe the $complex$ of the complex networks.
\section{Application}
\label{application}
In this section, the Tsallis entropy of the complex networks is used in these real networks, such as the the US-airlines networks(US-Airlines) \cite{networkdata}, Email networks(Email-network) \cite{networkdata}, the Germany highway networks(GM-Highway) \cite{nettt} and the protein-protein interaction network in budding yeast(Yeast-protein) \cite{networkdata}. The results are shown as follows.

\begin{table}[htbp]
\tiny
\addtolength{\tabcolsep}{-3pt}
  \centering
  \caption{The Tsallis entropy of these real networks with different value of $q$}
    \begin{tabular}{ccccccccccc}
    \toprule
        The value of $q$  & $q$=0.5 & $q$=0.7 & $q$=0.9 & $q$=1   & $q$=1.1 & $q$=1.3 & $q$=1.5 & $q$=1.7 & $q$=1.9 & $q$=2.1 \\
    \midrule
    US-Airlines \cite{networkdata} & 27.57983 & 12.6853 & 6.637035 & 5.025024 & 3.912498 & 2.555764 & 1.814395 & 1.37422 & 1.093357 & 0.90287 \\
    Email-network \cite{networkdata} & 58.2606 & 21.80083 & 9.471894 & 6.631019 & 4.831283 & 2.864851 & 1.921935 & 1.412793 & 1.10758 & 0.908247 \\
    GM-Highway \cite{nettt}  & 65.29844 & 24.03428 & 10.14312 & 6.994691 & 5.027314 & 2.921232 & 1.938028 & 1.417374 & 1.108885 & 0.90862 \\
    Yeast-protein \cite{networkdata}  & 78.90924 & 26.06444 & 10.37936 & 7.053887 & 5.029496 & 2.909061 & 1.931949 & 1.415083 & 1.108105 & 0.908367 \\
    \bottomrule
    \end{tabular}%
  \label{tab:addlabel}%
\end{table}%

\begin{figure}[htbp]
  \centering
  \subfigure[US-Airlines ]{
    \label{Local_networkSS:subfig:b} 
    \centering
    \includegraphics[scale=0.42]{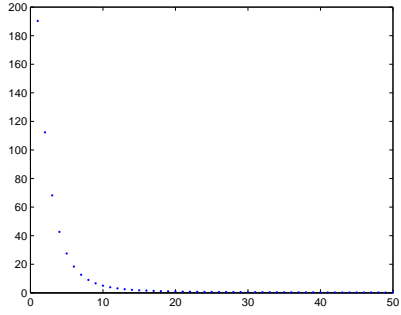}}
  \hspace{0.5cm}
    \subfigure[Email-network]{
    \label{Local_networkSS:subfig:b} 
    \centering
    \includegraphics[scale=0.42]{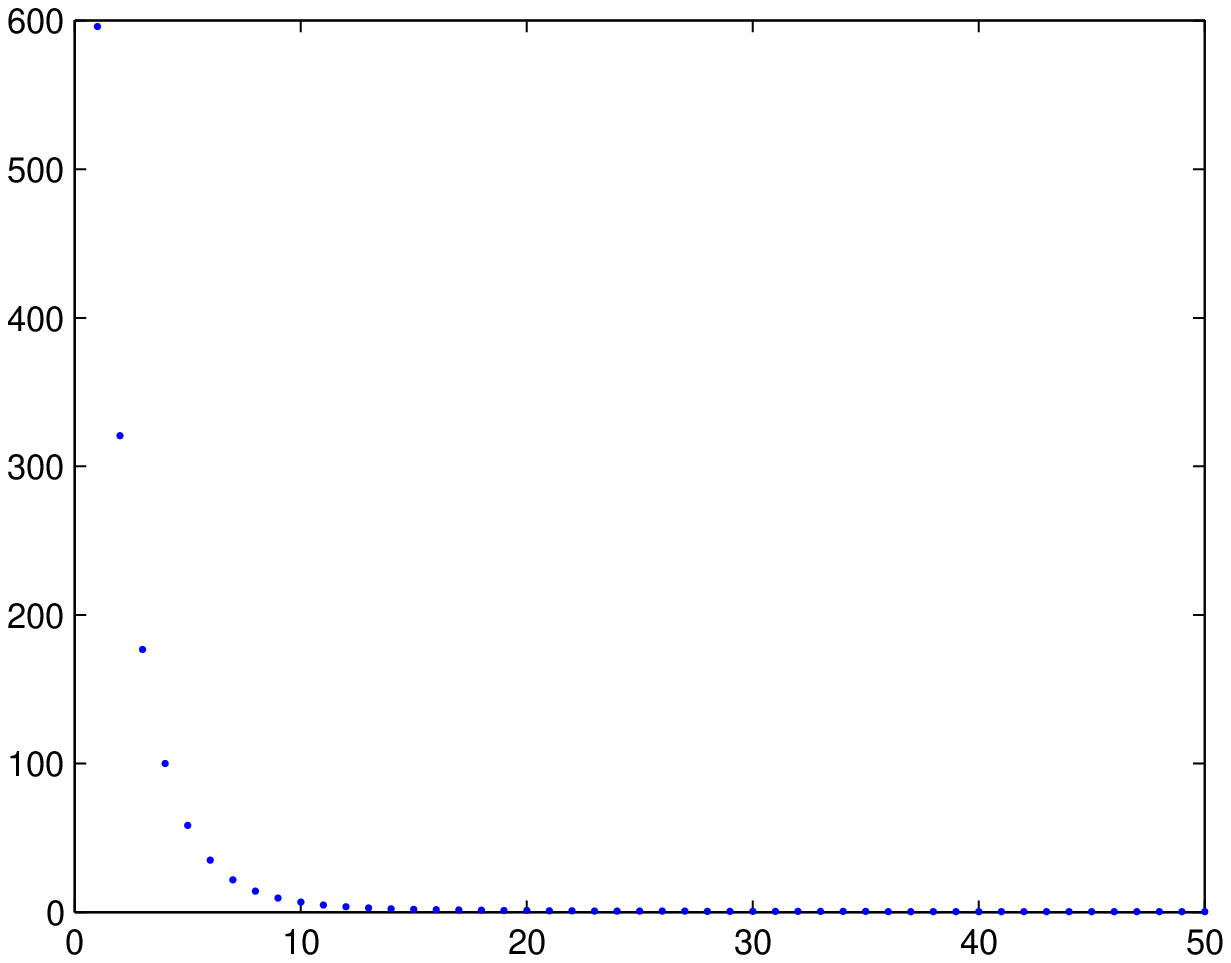}}
  \hspace{0.5cm}
    \subfigure[GM-Highway]{
    \label{Local_networkSS:subfig:c} 
    \centering
    \includegraphics[scale=0.42]{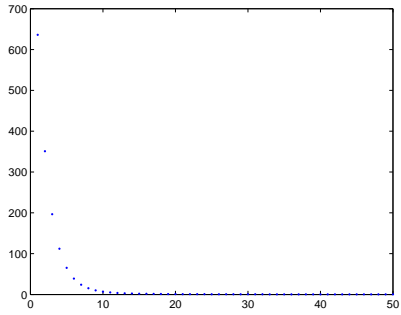}}
  \hspace{0.5cm}
      \subfigure[Yeast-protein]{
    \label{Local_networkSS:subfig:d} 
    \centering
    \includegraphics[scale=0.42]{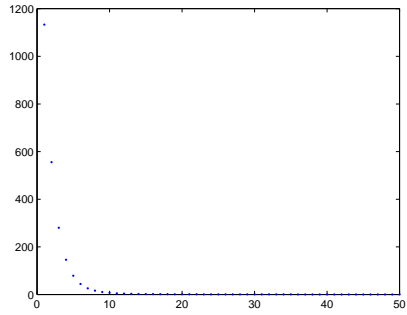}}
  \caption{The abscissa in these subfigure represents value of $Step$, the value of $q$ is decided by the value of $step$, $(q=Step\times0.1)$. The borders of $q$ in this simulation is $[0,5]$. The ordinate in these subfigure represents the value of the entropy. The point in these subfigure represents the Tsallis entropy of each real network with different value of $q$. The details in these subfigures show that follow the increase of the value of $q$ the value of the Tsallis entropy for each real network is equal to 0.  }\label{Local_networkSS}
\end{figure}

The results in the application of the Tsallis entropy show that the Tsallis entropy can be used to describe the $complex$ of the complex networks. The value of the Tsallis entropy of the complex network is corresponded to the scale of the complex networks. The bigger the scale of the complex networks, the more the $complex$ of the complex networks.
\section{Conclusion}
\label{conclusion}
In the existing research on the $complex$ of the complex networks, the main idea is that the structure of the complex networks decide the degree of the $complex$ in the complex networks. However, the Tsallis entropy of the complex networks which is proposed in this paper show that the $complex$ of the complex networks also is decided by the relationship between the nodes. Chose different kinds of nodes as the majority of the network will influence the $complex$ of the complex networks. The value of $q$ in the proposed method decided which kind of nodes are chosen as the main part of the network. The Tsallis entropy of the complex networks have expanded the method on the description of the $complex$  of the complex networks. It is an generalise method to describe the property of the complex networks.

\section*{Acknowledgments}
The work is partially supported by National Natural Science Foundation of China (Grant No. 61174022), Specialized Research Fund for the Doctoral Program of Higher Education (Grant No. 20131102130002), R$\&$D Program of China (2012BAH07B01), National High Technology Research and Development Program of China (863 Program) (Grant No. 2013AA013801), the open funding project of State Key Laboratory of Virtual Reality Technology and Systems, Beihang University (Grant No.BUAA-VR-14KF-02).

%



\bibliographystyle{elsarticle-num}
\bibliography{zqreference}






\end{document}